\documentclass[multphys,vecphys]{svmult}

\usepackage{makeidx}         
\usepackage{graphicx}        
\usepackage{multicol}        
\usepackage[bottom]{footmisc}

\usepackage{amsmath}
\usepackage{amssymb}
\makeindex             

\begin{document}

\title*{VITRUV Science Cases}
\author{                                 Paulo~J.~V.~Garcia\inst{1}\and
  Jean-Phillipe~Berger\inst{2}\and           Romano~Corradi\inst{3}\and
  Thierry~Forveille\inst{4}\and                 Tim~Harries\inst{5}\and
  Gilles~Henri\inst{2}\and                    Fabien~Malbet\inst{2}\and
  Alessandro~Marconi\inst{6}\and             Karine~Perraut\inst{2}\and
  Pierre-Olivier~Petrucci\inst{2}\and       Karel~Schrijver\inst{7}\and
  Leonardo~Testi\inst{6}\and                  Eric~Thiebaut\inst{8}\and
  Sebastian~Wolf\inst{9}}

\institute{ Departamento de F\'{\i}sica  da Faculdade de Engenharia \&
  Centro   de  Astrof\'{\i}sica,   Universidade  do   Porto,  Portugal
  \texttt{pgarcia@astro.up.pt}\and   Laboratoire  d'Astrophysique  UMR
  UJF-CNRS 5571, Universit\'e  Joseph Fourier, France\and Isaac Newton
  Group   of  Telescopes,  Spain\and   Canada-France-Hawaii  Telescope
  Corporation,  Hawaii,  USA\and  School  of  Physics,  University  of
  Exeter, UK\and INAF - Osservatorio Astrofisico di Arcetri, Italy\and
  Lockheed  Martin  Advanced  Technology Center,  California,  USA\and
  Observatoire  de  Lyon/CRAL,  France\and  Max Planck  Institute  for
  Astronomy - Heidelberg, Germany}
%
%
\authorrunning{The Vitruv Science Group}

\maketitle

\begin{abstract}
  
  VITRUV is  a second generation spectro-imager for  the PRIMA enabled
  Very Large Telescope  Interferometer. By combining simultaneously up
  to 8 telescopes VITRUV makes the  VLTI up to 6 times more efficient. 
  This operational gain allows  two novel scientific methodologies: 1)
  massive surveys  of sizes; 2) routine  interferometric imaging.  The
  science cases presented concentrate on the qualitatively new routine
  interferometric  imaging  methodology.  The  science  cases are  not
  exhaustive  but complementary  to the  PRIMA reference  mission. The
  focus is on: a) the close  environment of young stars probing for the
  initial conditions  of planet formation  and disk evolution;  b) the
  surfaces  of stars tackling  dynamos, activity,  pulsation, mass-loss
  and  evolution;   c)  revealing  the  origin   of  the  extraordinary
  morphologies of Planetary Nebulae and related stars; d) studying the
  accretion-ejection structures  of stellar black-holes (microquasars)
  in our  galaxy; e)  unveiling the different  interacting components
  (torus, jets,  BLRs) of Active  Galactic Nuclei; and f)  probing the
  environment  of  nearby  supermassive black-holes  and  relativistic
  effects in the Galactic Center black-hole.

\end{abstract}


\section{Introduction}

The  idea behind VITRUV  is that  by combining  simultaneously 4  to 8
telescopes the VLTI becomes 2 to  6 times more efficient than by using
non-simultaneously  these  same  telescopes.  Therefore  an  immediate
operational gain  is achieved and two  scientific methodologies become
possible:  1) massive  surveys  of sizes;  2) routine  interferometric
imaging.  The science cases presented here are essentially centered on
routine interferometric  imaging as this is the  novel VLTI capability
made possible by VITRUV.

\subsection{Imaging with the VLTI}

\begin{figure}[h]
  \begin{center}
    \resizebox{\textwidth}{!}{\includegraphics{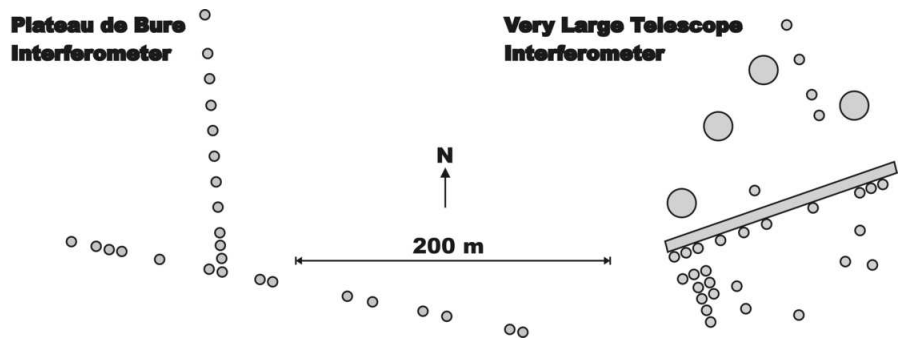}}
  \end{center}
  \caption{The Plateau  de Bure  Interferometer (PdBI) and  Very Large
    Telescope Interferometer  (VLTI) have similar  baselines.  Because
    the VLTI operates at $\sim 1 \mu$m and the PdBI at $\sim 1$~mm the
    angular resolution of the VLTI  is $\sim 1$~milliarcsec and of the
    PdBI $\sim 1$~arcsec.\label{garcia_f1}}
\end{figure}

With the full operation of the  AMBER instrument and 3 movable ATs the
Paranal  Observatory  will   have  the  same  interferometric  imaging
capabilities of  the Plateau  de Bure Observatory  at it's  dawn.  The
VLTI  and  PdBI  have  similar baselines  (Fig.~\ref{garcia_f1}),  but
because the  VLTI operates  at a wavelength  up to 1000  times shorter
than the PdBI its angular resolution will be up to 1000 times sharper.

In  interferometry, imaging  is achieved  by inverting  a sufficiently
large  number  of complex  visibilities.   These  can  be obtained  by
triangular combinations of telescopes.  A sufficiently large number of
complex visibilities,  typically six different  antenna triangles, were
used  by the  PdBI together  with  a significant  integration in  hour
angle.  As  the hour angle  changes, the triangular projection  in the
sky  rotates.   For  a   given  triangle,  many  more  (8-16)  complex
visibilities  are  available  for  large  (-4~h  to  4~h)  hour  angle
variations.  Imaging  with 3  telescopes is therefore  time consuming.
One image requires a complete night for each of the 6 triangles, i.e.,
six nights are required to image one object.  This was common practice
at  the PdBI  in  its early  days --  a  typical example  is given  by
Guilloteau et al.(1992).

The   conclusion  to   draw   from  the   PdBI   experience  is   that
interferometric imaging  of compact sources with a  quality similar to
the PdBI with  AMBER and the ATs is possible but  it will be expensive
in terms of telescope time and an operational burden.

\subsection{The VITRUV instrument}

VITRUV is a general purpose spectro-imager for the PRIMA enabled VLTI.
The instrument is  described in detail in these  proceedings by Malbet
et al.  (2005) In the following  sections we concentrate in  a few key
areas  where  VITRUV  spectro-imaging  will make  a  significant  step
forward.   These  areas are  not  exhaustive  and  should be  seen  as
complementary to the PRIMA reference mission document.

VITRUV  being   a  spectro-imager   has  three  modes   with  spectral
resolutions  of 100,  1000  and 10000.   These  modes are  essentially
dictated by  the science cases: low resolution  for continuum imaging,
intermediate resolution for line  emission imaging and high resolution
for stellar surface imaging and line emission kinematics. The spectral
coverage  is centered  on JHK  with possibilities  of extending  it to
visible (R+I) and  L bands.  The visible region  allows higher angular
resolution  and access  to  scientifically interesting  lines such  as
H$\alpha$  and  the  Calcium  triplet.   The L  band  extension  allow
polycyclic aromatic hydrocarbons (PAHs) and nanodiamonds science.


\section{The formation of stars and planets}

Accretion and  outflow are the distinctive  simultaneous signatures of
star formation.   However even for the  best studied case  of {\bf low
mass stars ($M  \lesssim 2 M_\odot$)} the physical  mechanism by which
matter is  accreted and ejected  remains unknown.  The  angular scales
responsible for  the bulk of  the observed SED emission  (specially in
the  optical-NIR) and the  angular scales  where the  outflow activity
originates  are smaller  than those  probed by  adaptive optics  on 8m
class telescopes -- any progress  is only possible by directly imaging
the optical/NIR emission lines and continuum at $1'''$ resolution.

Only  recently  was the  existence  of  Keplerian  gas disks  in  {\bf
  intermediate mass stars ($2 M_\odot \lesssim M \lesssim 8 M_\odot$)}
demonstrated  with mm~interferometric  imaging  (Mannings \&  Sargent,
1997).  NIR-interferometric observations have shown (e.g.  Millan-Gabet
et al., 2001) that the bulk of  the NIR emission comes not from a disk
but from from  a puffed-up inner wall located  at the dust sublimation
radius.  This completely different disk geometry from their lower mass
analogs  underlines how imaging  information is  critical for  the SED
understanding.

{\bf  High mass  stars  ($M\gtrsim  8 M_\odot$)}  have  a very  strong
radiation  field that  photoionizes  the surrounding  and  also has  a
dynamical effect via radiation  pressure.  Powerful mass ejection also
takes place.  Only around  ten disk-like structures have been detected
so  far and it  is not  clear if  some of  them are  actually unstable
infalling  material  rather  than  Keplerian  disks.   Recently  2MASS
counterparts have  been found around a sample  of massive protostellar
candidates (Kumar  et al.  2005), opening the  exciting possibility of
studying    the   close    environment   of    these    objects   with
NIR-interferometry.

\subsection{The structure of inner disks of young stars}

The  morphology of  the dusty  close environment  of pre-main-sequence
stars  will   be  studied   by  VITRUV,  particularly   the  following
interconnected aspects:\\ {\bf a)} What are the effects of the central
radiation field  in the  environment structure --  exact shape  of the
sublimation  surface/rim and  inner cavity?\\  {\bf b)}  How  does the
morphology  correlates  with dust  properties?\\  {\bf  c)} Are  there
orbiting companions opening cavities in the central disk region?  What
are  consequences on  companion formation  and  migration scenarios?\\
{\bf  d)}  Is the  environment  shaped  by  central source  winds  and
outflows?  Are these winds dusty?\\ {\bf e)} How does the structure of
the inner disk affects  the initial conditions for planet formation?\\
{\bf  f)}  What  is  the  distribution  and  morphology  of  PAHs  and
nanodiamonds in the inner regions of disks?

\noindent
{\bf Time dependent  morphology} At Taurus the Earth  orbit is located
at  7 mas.   For the  first  time a  systematic study  of the  orbital
evolution of  the dusty environment  will be feasible and  compared to
hydrocode simulations.

\noindent
{\bf   Evolutionary  aspects}  By   comparing  objects   at  different
evolutionary stages the timescales for the morphological evolution and
dissipation can  be addressed.  With  AMBER a considerable  advance in
PMS stellar  evolution models is  expected and more precise  timing of
the central stars will be available by 2010.

\noindent
{\bf Central  source mass}  The stellar mass  will affect  the central
radiation field and the  dusty environment via sublimation/heating and
pressure.

\begin{figure}[h]
  \begin{center}
    \resizebox{!}{3.6cm}{\includegraphics{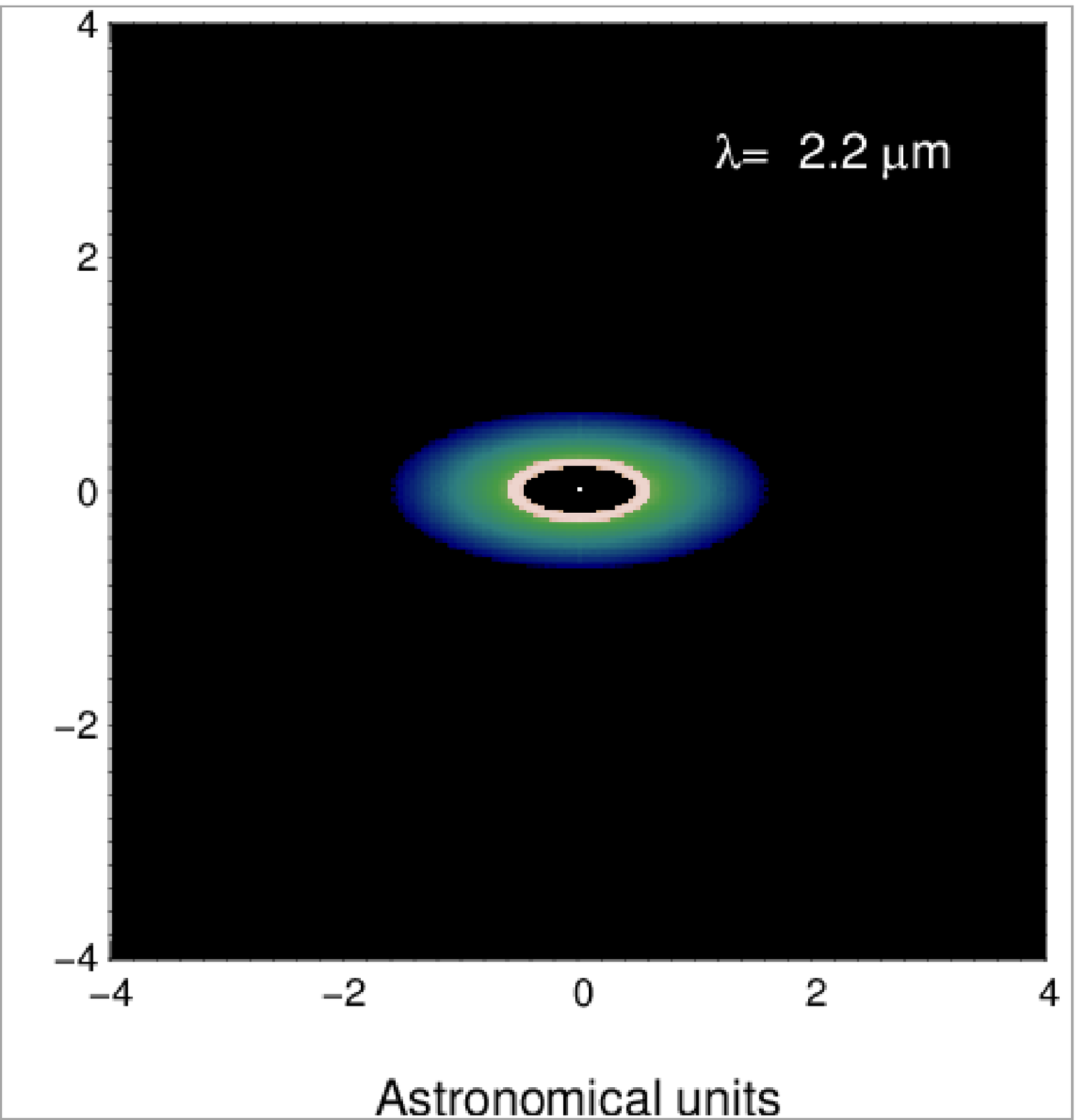}}
    \hspace{1cm}
    \resizebox{!}{3.6cm}{\includegraphics{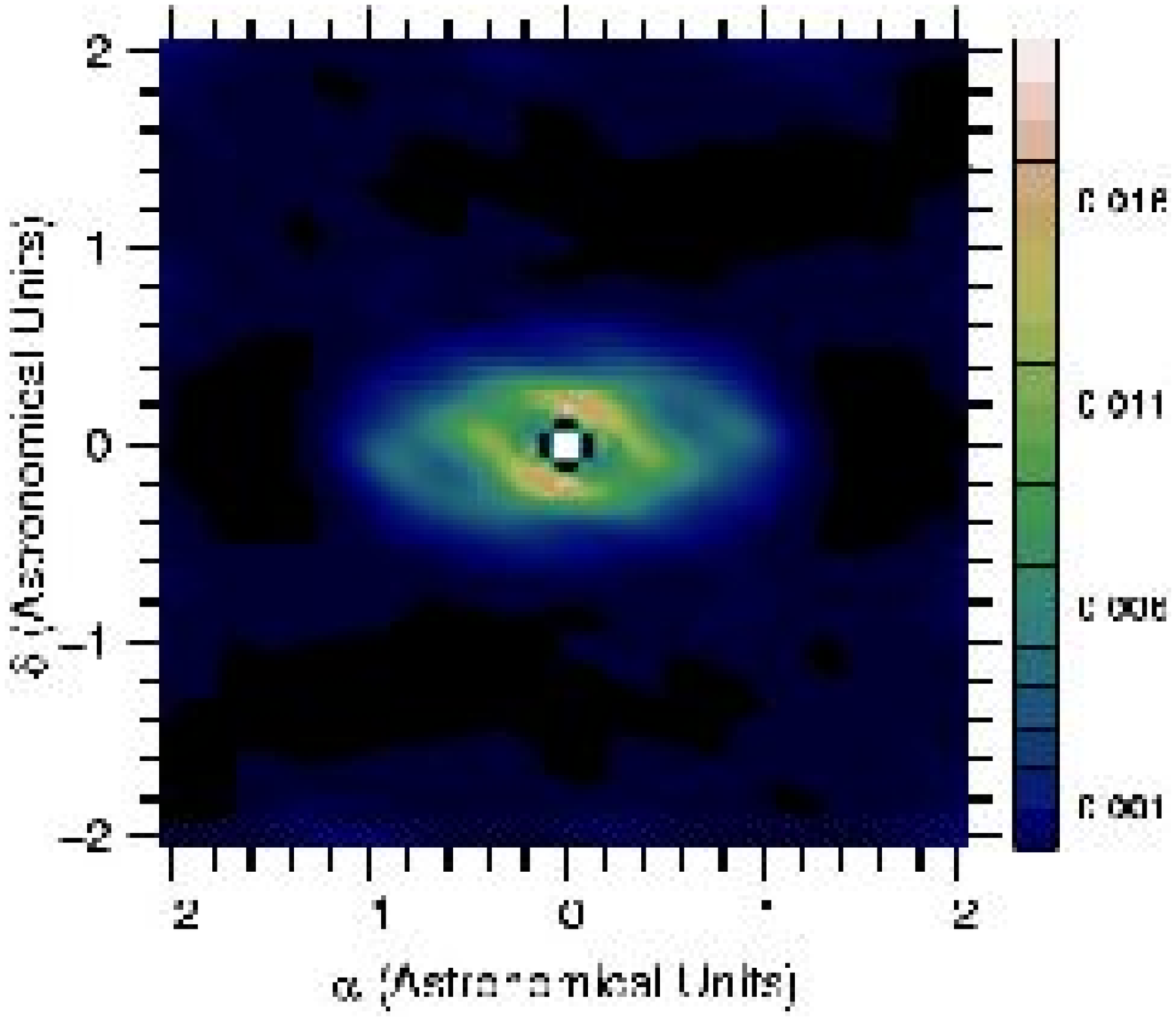}}
  \end{center}
  \caption{\label{disk_J}  Left:  puffed-up  inner  wall  model  (star
  removed). Right:  reconstructed image Thi\'ebaut  \& Tatulli (including
  central star). The ring is 15 times fainter than the central star.}
\end{figure}

\subsection{The launching of jets and winds}

The morphology  of the  emission lines in  close environment  of young
stars  will  be  studied,  particularly the  following  interconnected
aspects:\\ {\bf a)} Where does  the jet originates, in the disk, star,
or  disk-star interaction  region?\\ {\bf  b)} What  is  the mechanism
responsible  for maintaining  the  ionization in  the outer  launching
regions  as required  by  MHD models?\\  {\bf  c)} How  is the  strong
magnetic field  required for the jet  launching at the  wind base made
compatible   with   the   equipartition   regime   required   by   the
magneto-rotational instability?\\  {\bf d)} Energetically  the magnetic
engine is  a way  to tap gravitational  accretion power from  the disk
into the jet.  How can this made compatible with the SED modeling that
taps the  same energy in the  form of viscous  dissipation?\\ {\bf e)}
What  is the  relation between  the observed  dust morphology  and the
jets/winds?\\  {\bf f)}  What are  the chemical,  ionization, thermal,
kinematic and morphological properties (collimation, wiggling) of jets
at their launching  region?\\ {\bf g)} What is  the exact contribution
of  accretion  and  winds  to  the observed  infrared  hydrogen  lines
emission profile?

\noindent
{\bf  Time  dependent  morphology}   Is  wind  and  jet  launching  an
intrinsically time dependent  phenomenon (related to some instability)
or  what the  time variability  is  simply a  consequence of  variable
accretion in a otherwise stationary engine?  Are there orbital effects
in the ejection?

\noindent
{\bf Evolutionary  aspects} How  do the jet  and winds  properties (in
particular collimation  and excitation)  evolve in time  and correlate
with disk accretion?

\noindent
{\bf Central source mass} How does the central jet engine evolves with
the central mass, in particular  what are the effects of different: 1)
radiation fields; 2) stellar magnetic fields; 3) disk structures.

\noindent
{\bf Radiative transfer} Two breakthroughs  are expected: a) 2D and 3D
radiative transfer  will be required  to analyze the  observations; b)
clumpiness is  expected as  the ejection process  is unstable  -- this
stochastic density  along the line  of sight will radically  alter the
radiative transfer.

\begin{figure}[h]
  \begin{center}
    \resizebox{!}{4cm}{\includegraphics{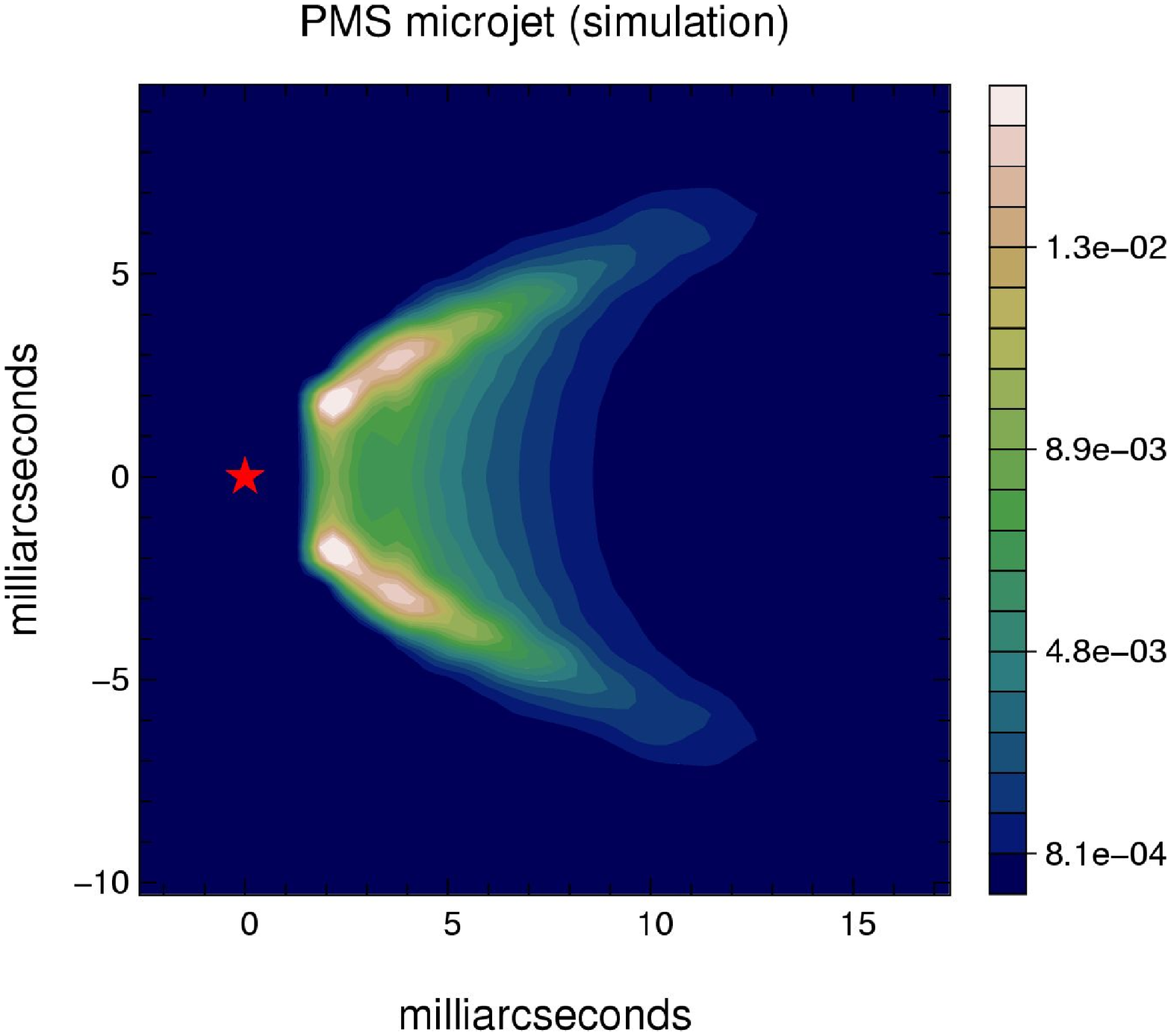}}
    \hspace{1cm}
    \resizebox{!}{4cm}{\includegraphics{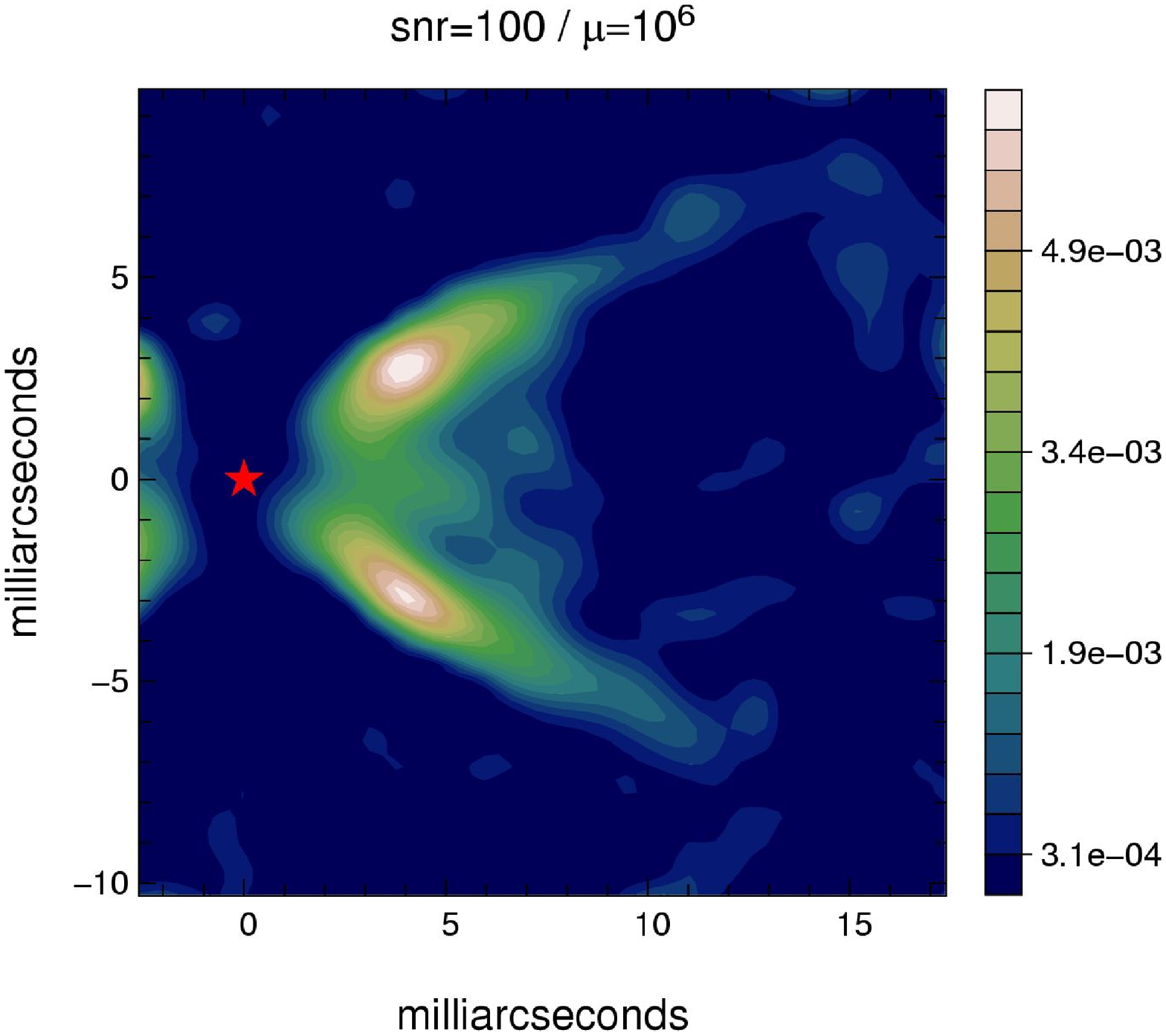}}
  \end{center}
\caption{Pa$\beta$ emission from the inner  region of a MHD disk wind,
 (left) and reconstructed  image by Thi\'ebaut et al  (right).  The star
 contribution, which is on average  has a surface brightness 100 times
 the jet has been removed.}
\end{figure}


\section{Imaging stellar surfaces}

VITRUV, as  an imaging device,  open the possibility to  study various
surface  properties as vertical  and horizontal  temperature profiles,
abundance  inhomogeneities  and detect  their  variability. These  key
observations  will  address   stellar  activity  processes,  mass-loss
events, magneto-hydrodynamic mechanisms, and stellar evolution.

\subsection{Photospheric convection and activity of late type stars}

{\bf Dynamos} The variable magnetic field that causes stellar activity
is  generated by  a process  that  we call  the dynamo,  in which  the
kinetic energy in convection and large-scale circulations is converted
into magnetic  energy.  Ab  initio, comprehensive modeling  of stellar
dynamos  is at  present  impossible; instead  modelers  are forced  to
introduce  multiple  simplifications  and  approximations  into  their
models,  without knowing  whether these  are warranted,  or  what side
effects  may be  introduced by  them. Consequently,  the  diversity of
results   from   currently   existing   models   critically   requires
observational  guidance  on,  for  example,  how  the  field-emergence
patterns change over periods of  years to decades. Repeated imaging of
stellar  chromospheric  emission  patterns  over multiple  years  with
VITRUV will provide the  essential observational knowledge on which of
these patterns actually  occur on other stars, and  how they depend on
fundamental stellar parameters as well as rotation rate.

\noindent
{\bf Spots on magnetically active  stars} Energy of late type stars is
mainly  carried outwards  by  convection whose  study  is critical  in
modeling  the inner  structure  as  well as  the  atmosphere of  these
objects.   VITRUV  imaging  of  the  turbulent eddies  of  such  stars
together with classical interferometric measurements of their size and
luminosity  (or  temperature) will  allow  a  better understanding  of
stellar convection.  Late type stars might exhibit asymmetric and even
highly fragmented mass-loss events, which are believed to be linked to
stellar   surface  parameters   such   as  limb-darkening,   effective
temperature or surface features.

\subsubsection*{Pulsating Miras}

\noindent
{\bf Surface temperature and  chemical inhomogeneity} VITRUV will test
chemically  inhomogeneous  mass-loss in  the  AGB  phase.  Of  further
interest is the suggested existence of magnetic cool spots produced by
a relatively weak large-scale asymmetric magnetic field amplified by
turbulent dynamo  on the surface of  Miras as the  cause of aspherical
mass distribution.

\noindent
{\bf   Polarization    aspects}   Narrow-band   photo-polarimetry   and
spectro-polarimetric  observations demonstrated  that  the polarization
magnitude  varies through the  molecular bands  of Miras  and post-AGB
stars (e.g.   Trammell et  al. 1994).   Such an affect  may be  due to
changes in albedo with optical depth in the extended atmosphere.  When
spatial  information is available  the interpretation  of polarization
data is much  more strongly constrained.  Interferometric polarization
measurements  of late-type giant  stars with  VITRUV will  provide new
insights into  their atmospheric structure  and the geometry  of their
circumstellar material,  and the  chemical and physical  properties of
the dust  in the photosphere  and the winds, constraining  theories of
mass-loss from red giant  stars and hydrodynamical models of planetary
nebulae formation.

\subsection{Abundance and magnetic field of Ap stars}

Chemically Peculiar A  and B stars (CP stars)  exhibit strong chemical
abundance inhomogeneities  of one or  more chemical elements,  such as
helium, silicon,  chromium, strontium, or europium,  and a large-scale
organization of their magnetic field that produces a typical signature
in circularly-polarized  spectra. CP stars represent a  major class of
the  known magnetic  stars in  the solar  neighborhood  and constitute
ideal targets  for studying how magnetic fields  affect other physical
processes  occurring in  stellar  atmospheres.  VITRUV  simultaneously
mapping of abundance distributions and magnetic fields will tackle the
fundamental question of  the origin of the magnetic  field in CP stars
(Moss,  2001).  Both  the  fossil and  the  core-dynamo theories  have
difficulty in  explaining all  the observed magnetic  characteristics. 
These  same maps  will  allow to  better  understand the  key role  of
magnetism in a) atmosphere  structuration; b) ion migration across the
stellar surface and; c) chemical stratification.


\section{Evolved stars}

\subsection{The shaping of the outflows from evolved stars}

In  the last  twenty years,  the extraordinary  geometry of  young and
evolved  PNe and  related objects  (like for  instance  nebulae around
symbiotic   stars)  has   been  revealed   by  ground-based   and  HST
observations. These  observations gave a strong  impulse to theorists,
and a plethora of theoretical explanations has been proposed. However,
none of them  could be fully tested, as in most  cases even HST cannot
resolve the spatial scale at which collimation occurs.

The  most popular  models to  explain the  onset of  asymmetry  in the
outflows involve: a) strong interactions between (anisotropic) AGB and
post-AGB winds producing  wind-heated and wind-blown expanding bubbles
(Balick  \& Frank  2002); b)  collimation by  several  processes (like
accretion disk winds interacting with  the stellar winds) in close and
wide interacting binaries;  c) magneto-hydrodynamical (MHD) shaping --
this latter scenario is also possibly related to enhanced rotation and
magnetic fields due to binary interactions.

\begin{figure}[h]
  \begin{center}
    \resizebox{6cm}{!}{\includegraphics{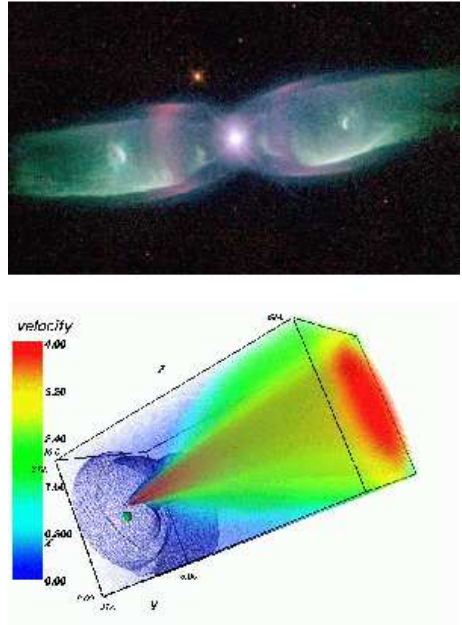}}
  \end{center}
  \caption{\label{m29}Top: HST image of  the bipolar nebula M 2-9. The
    long   side  of  the   f.o.v  is   about  30000~AU.   Bottom:  the
    hydrodynamical  simulation  by  Garcia-Arredondo \&  Frank  (2004)
    using  an accreting  interacting symbiotic-like  binary  with P=20
    yr.    The    model    box    includes    only    the    innermost
    320$\times$160$\times$160~AU.}
\end{figure}

\noindent
{\bf  Resolving the  collimation  region} In  many  of these  objects,
VITRUV will allow us to see directly the collimation region and follow
in real-time  its dynamical evolution,  revealing which one(s)  of the
models above  is correct.  Figure.~\ref{m29}, top,  presents the image
of the  highly collimated nebula  M~2-9 (d=640 pc), obtained  with the
HST.  At the bottom, on a  100 times smaller scale it is instead shown
the  hydrodynamical simulation  by Garcia-Arredondo  \&  Frank (2004),
where the  cell size of of 0.5~AU  and FoV matches those  of VITRUV at
the distance  of M~2-9.  At such  resolution, the 10~AU  offset of the
jet origin (produced by the unresolved hot companion and its accretion
disk)  with  respect  to   the  primary  star,  the  corkscrew  matter
distribution around the  stars, and the jet aperture  and bending, can
be clearly revealed and  would unambiguously demonstrate the nature of
the binary  collimation source.  VITRUV  will follow in  real-time the
evolution  of  the  outflows,   distinguish  among  all  the  proposed
scenarios,  as  each  model  predicts its  own  characteristic  growth
behavior.  By  measuring the  Doppler shifts of  lines emitted  by the
circumstellar nebulae will result  in detailed spatio-kinematical maps. 
Finally, physical conditions in  the circumstellar gas will be derived
from the emission line spectrum.

The same  scientific objectives described above also  apply to related
objects, like the outflows from  symbiotic novae, classical novae , as
well as to nebulae around more massive stars, like LBVs.

\noindent
{\bf  Expansion  parallaxes} The  ability  of  resolving the  apparent
expansion  of the  nebulae, while  measuring  at the  same time  their
radial velocities, will also allow us to determine the distance of the
nebulae via their expansion parallax.   This is one of the most robust
methods to  determine the distances  of these objects  which represent
the  most  severe  limitation  to  determine  their  basic  physical
properties, like the luminosity, mass, age, and energetics.

\noindent
{\bf Binarity} In a number of  cases where the central regions are not
severely ``polluted'' by a large amount of gas and dust, VITRUV imaging
will  be able  to detect  and resolve  the two  stellar  components in
binary systems.   This would be another important  result, as binarity
effects  might  explain most  of  (if  not  all) the  deviations  from
spherical  symmetry  observed in  the  outflows  from  evolved stars.  
According to the theoretical models  two third of all PN central stars
are binaries,  the great majority  of which have orbital  periods much
larger  of 1~AU,  resolved by  VITRUV.  The  VITRUV detection  rate of
binaries is  expected to be  much higher and sampling  a complementary
separation space  than previous  surveys. For objects  at intermediate
separations, orbital  parameters will also be  obtained by multi-epoch
observations.

\subsection{Microquasars (stellar black-holes)}

Microquasars (Mirabel \& Rodr\'{\i}guez 1994), can be seen as galactic
counterparts of the extragalactic quasars, but on a much smaller scale
since they are thought to harbor  stellar mass black holes ($M \sim 10
M_\odot$).   Radio   images  of  microquasars   show  either  extended
structures like in  SS433 or very compact ones like in  Cyg X-1 at mas
scale and, in  general, radio emission is interpreted  as the presence
of  compact steady  jets  or  sporadic ejection  events  also seen  in
infrared and X-ray bands.  Due  to their smaller physical size compare
to AGNs, their radio emission  can be strongly variable and correlated
to  various  X-ray  states,  which  indicates  that  jet  emission  is
physically linked  to the  accretion process. Radio  observations have
also shown pairs of bright radio knots moving at apparent superluminal
velocities.

\begin{figure}[h]
  \begin{center}
    \resizebox{!}{3cm}{\includegraphics{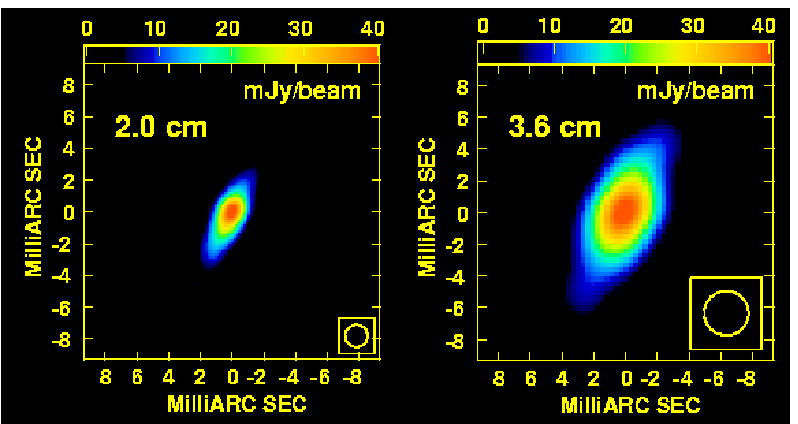}}
    \hspace{1cm}
    \resizebox{!}{3.5cm}{\includegraphics{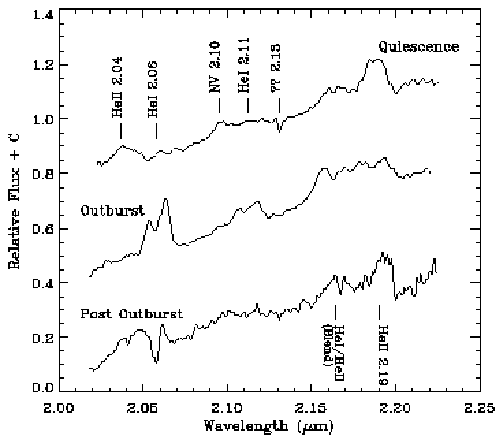}}
    \caption{Left:  VLBA images at  2.0 and  3.6 cm  on April  2, 2003
    showing the  compact jet.   The convolving beams  are 1.4  and 2.8
    mas, respectively.  1 mas corresponds to 12 AU at 12 kpc distance.
    (Fuchs  et  al. 2003)  Right:  K-band  spectra  of Cyg  X-3,  each
    displayed with an arbitrary flux offset for clarity.  At the top a
    quiescent spectrum, in  the middle a spectrum taken  during a very
    extreme  radio/X-ray  flaring  of  the  system, at  the  bottom  a
    "post-outburst." spectrum (Hanson et al. 2000).}
  \end{center}
\end{figure}

The light curves of microquasars in all wave-band exhibit very complex
time  behavior  connected to  the  accretion-ejection mechanism.   The
origin  of the  infrared emission  in microquasars  is far  from being
completely understood. Open questions are: a) What is the dust and jet
contribution  to the  infrared emission  of the  innermost  regions of
microquasars? b) How  does it connect with the  radio jet emission? c)
How does it vary with the X-ray spectral states?

Infrared imaging of  microquasars at mas scale has  never been done up
to now.   K band images of  microquasars at mas  resolution will allow
to: a) study the jet  morphology; b) constrain the different parameter
distributions  (particle distribution,  density,  magnetic field)  all
along the jet; c) constrain accretion-ejection models; d) characterize
dust in the inner region of these objects.


\section{Active Galactic Nuclei}

Active Galactic Nuclei (hereafter  AGN) are galactic nuclei powered by
non-stellar  energy production.  Marconi et  al. (2001)  summarize the
expectations from  AGN observations with the first  generation of VLTI
instruments  (AMBER  and MIDI).   Although  very important  scientific
results can be achieved with AMBER  and MIDI (e.g., Jaffe et al. 2004),
it  is clear  that without  any imaging  capability,  observations are
limited  only to  the estimate  of source  sizes.  A  significant step
forward  with  respect to  AMBER  and MIDI  can  only  be obtained  by
providing images  of the  circumnuclear environment of  an AGN  at mas
resolution  (i.e.~$\sim$ sub-pc  scales).  In  the following,  we will
show  the possibilities of  the combined  use of  VITRUV and  PRIMA in
addressing several open issues on AGNs and supermassive Black Holes.

\begin{figure}
  \centering
  \includegraphics[width=0.48\textwidth]{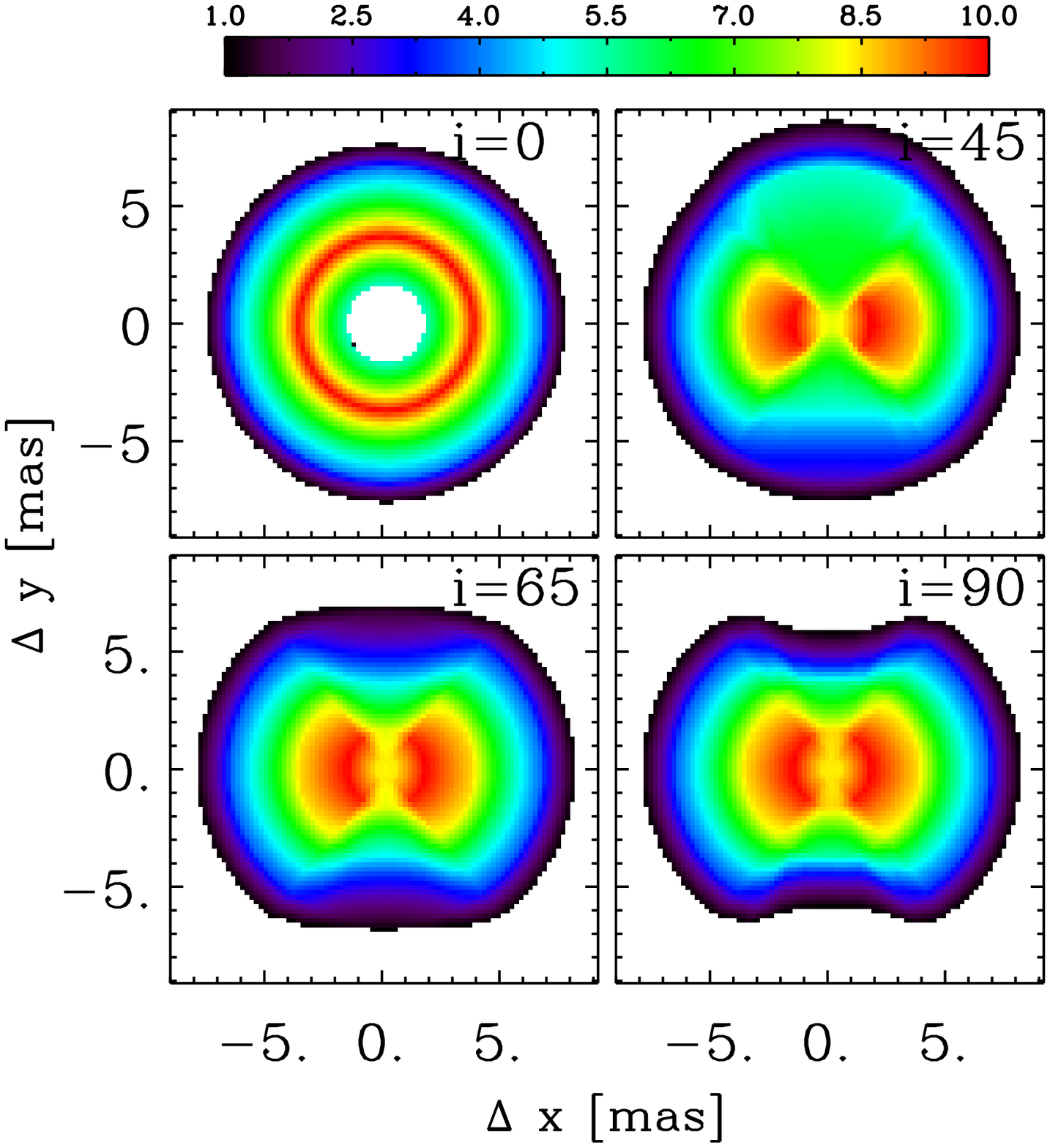}
  \includegraphics[width=0.48\textwidth]{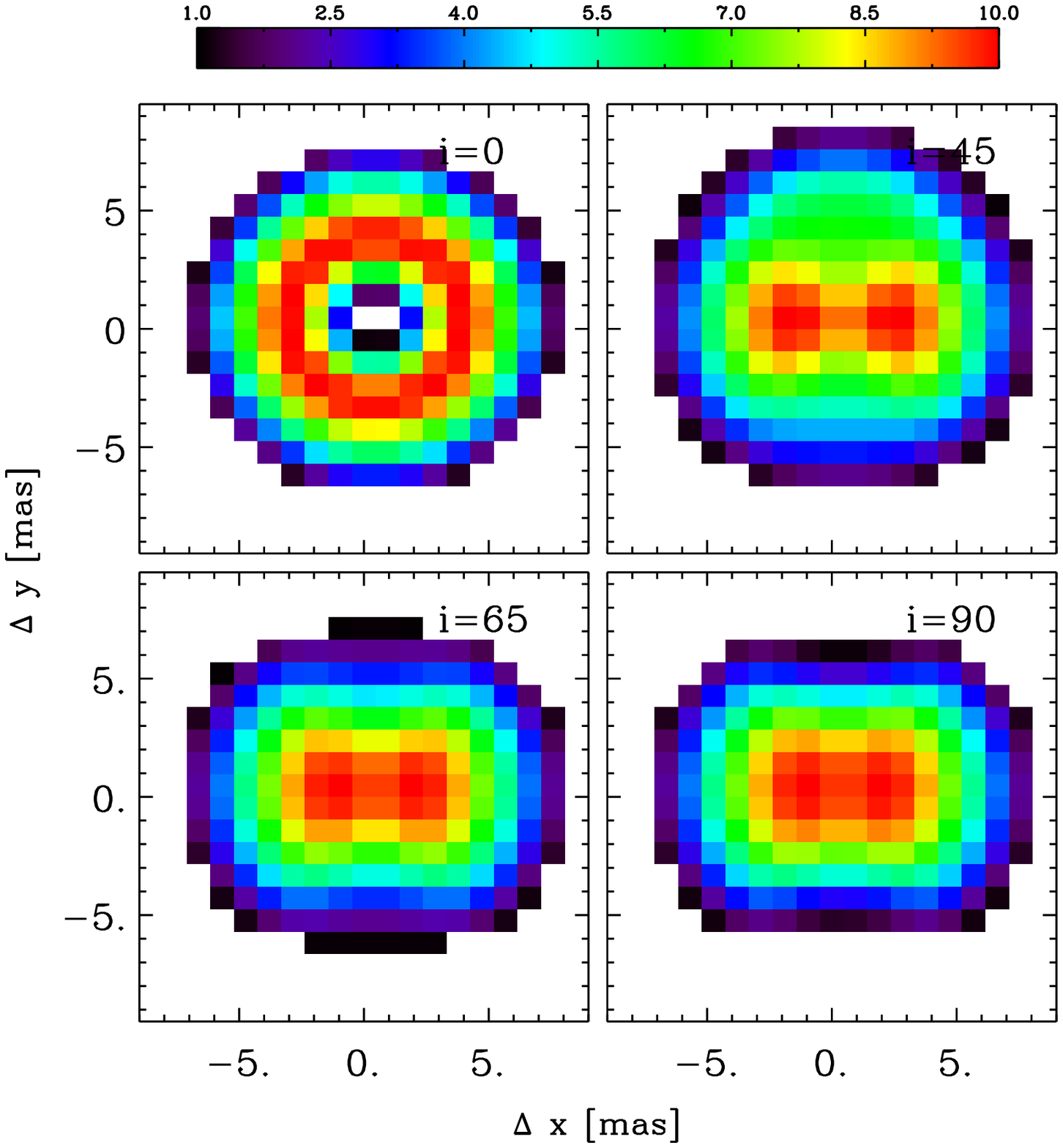}
  \caption{\label{fig:torus} Left:  Model images  of the NGC  1068 torus
    (Granato et  al.  2004)  seen at different  inclinations in  the K
    band. $i$ is  the inclination of the torus  pole axis with respect
    to  the line of  sight (with  $i=0$ the  torus is  seen face  on). 
    Images  are   in  units  of  $10^{-10}$   erg  s$^{-1}$  cm$^{-2}$
    \AA$^{-1}$  except  for the  $i=0$  image  which  is in  units  of
    $10^{-10}$ erg  s$^{-1}$ cm$^{-2}$ \AA$^{-1}$.  Right: same images
    as before but convolved with a PSF with 2 mas FWHM and re-binned to
    1 mas pixels.}
\end{figure}

{\bf The  dusty torus} The dusty  torus is a  fundamental component of
the AGN unified model, but we do not know if it really exists and what
is  its real  morphology.  In  Fig.~\ref{fig:torus} we  present K-band
model images  of the  torus of NGC~1068  illustrating how VITRUV  K band
imaging of  the nuclear  region of  an AGN will  allow to:\\  {\bf a)}
confirm/negate   the  existence  of   tori;\\  {\bf   b)}  disentangle
synchrotron from  hot dust emission, i.e.   discriminate between torus
and jet emission;\\ {\bf c)}  study the torus morphology and determine
its geometrical  parameters like the  inclination w.r.t.  the  line of
sight;\\  {\bf d)}  constrain  radiative transfer  models;\\ {\bf  e)}
study the torus dust composition (using radiative transfer models).

\begin{figure}
  \centering
  \includegraphics[width=0.48\textwidth]{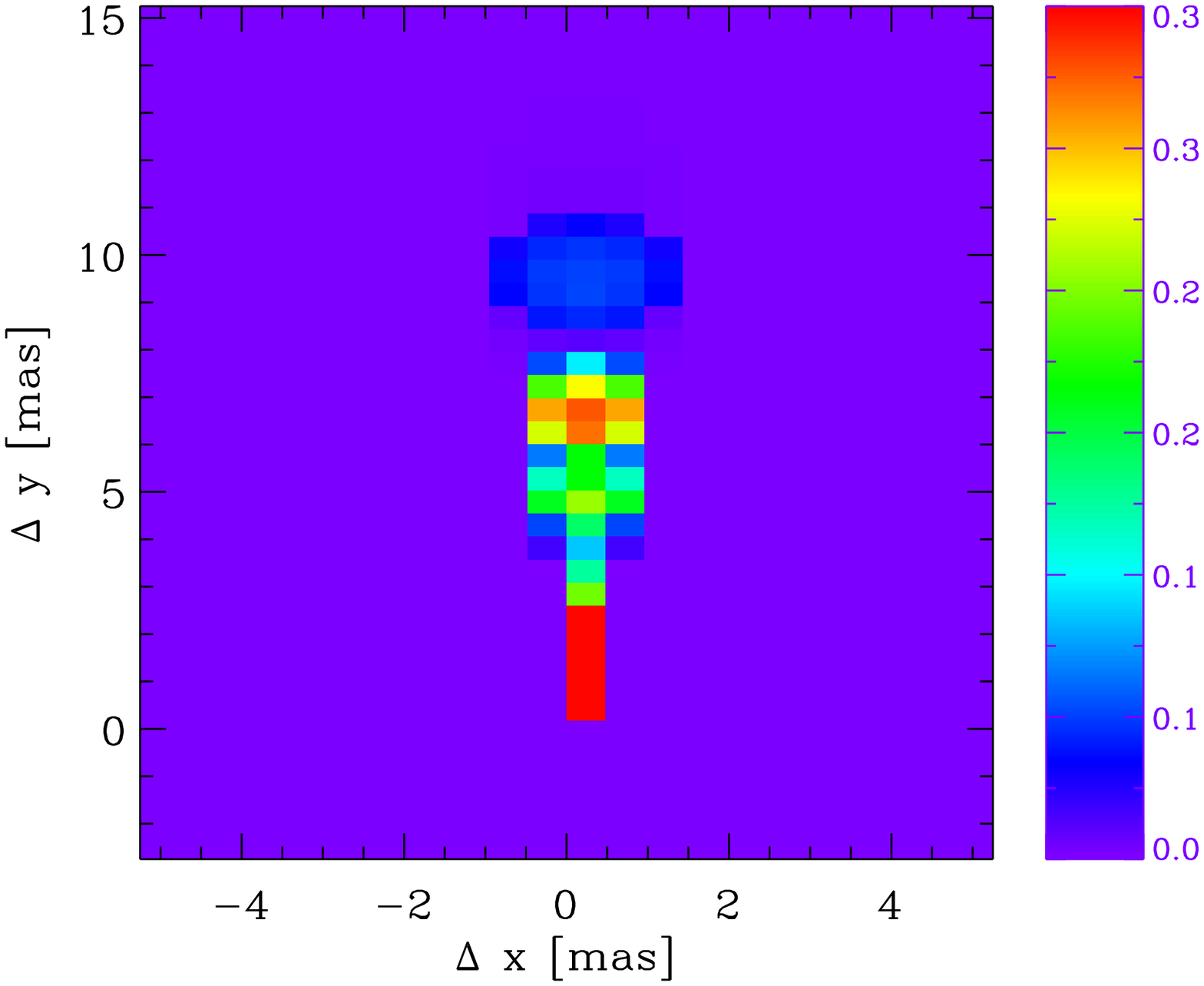}
  \includegraphics[width=0.48\textwidth]{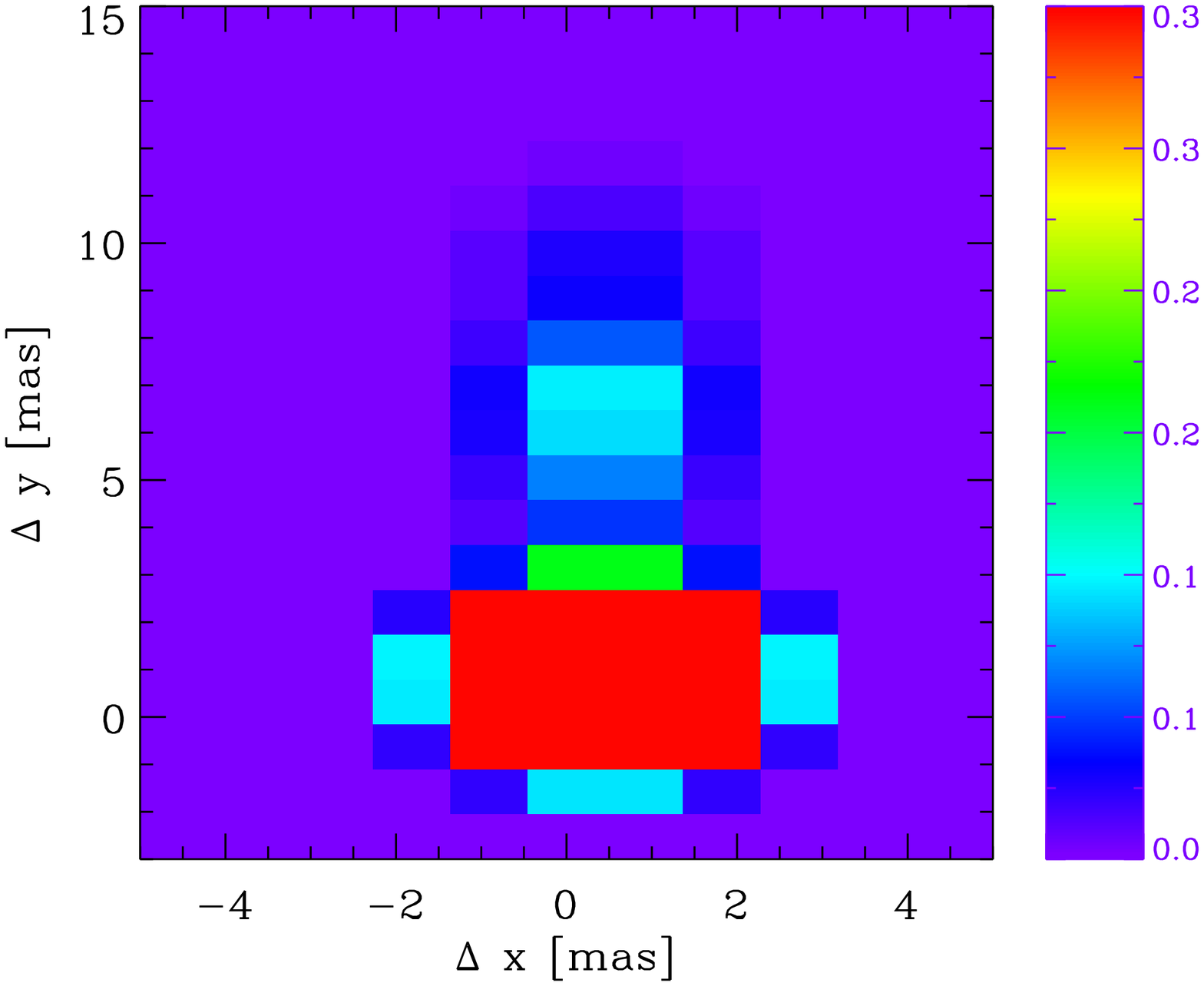}
  \caption{\label{fig:jet} Left: Model image of  the jet of 3C273 in the
    K band.  Image  is in units of mJy.  This  image assumed a conical
    jet with a  simple power law distribution for  the particles.  The
    density and  magnetic field distributions  along the jet  axis are
    tuned to reproduced real VLA  radio observations (Mantovani et al. 
    1999). Right: same model image as before but with a 2 mass spatial
    resolution and pixels of 1 mas.}
\end{figure}

{\bf  The Jet} Some  Active Galactic  Nuclei (AGNs)  exhibits powerful
collimated  jets.   Although a  minority,  they  are very  interesting
because, first  the ejection phenomenon from a  Black Hole environment
is very intriguing,  and also because jets are  sources of high energy
processes.  We present in  Fig.~\ref{fig:jet} a K-band model images of
the  jet of 3C  273. It  is worth  noting that,  in the  infrared, the
synchrotron self  absorption becomes important a  distance much closer
to the  central engine  than at radio  wavelength.  We will  then have
access for  the first time to  the jet inner  regions. Moreover, given
the broad-band (radio  up to X-rays) jet spectra  observed in AGNs, we
expect a high  energy cut off of the  synchrotron emission, which must
lie between radio  and optical.  Its detection is  important to better
constrain the maximal energy of  the emitting particle.  VITRUV K band
imaging of AGN jets at mas  resolution will allow to:\\ {\bf a)} study
the  jet  morphology;\\ {\bf  b)}  constrain  the different  parameter
distributions  (particle distribution,  density,  magnetic field)  all
along the  jet;\\ {\bf c)}  study the acceleration processes  that may
occur in the jet;\\ {\bf d)} constrain accretion-ejection models.

{\bf The  Broad Line  Region} The  BLR is the  region where  the broad
(FWHM$>1000$~km/s) permitted  lines observed in the spectra  of type 1
AGNs originate.   The morphology and  kinematics of the BLR  are still
unknown. The  spatial resolution  of VITRUV is  not enough  to provide
conventional  images of  the Broad  Line Region  however  by combining
medium  resolution spectroscopy  with accurate  phase  measurements it
will be  possible to  recover the photocenter  position of the  BLR in
each  wavelength (velocity)  bin.  This  will allow  to  constrain the
morphology and  kinematics of the BLR  it will be  possible to:\\ {\bf
  a)}  estimate   the  size  of   the  BLR  and  establish   a  secure
size-luminosity relation  for the BLR which is  fundamental for virial
mass estimates of BH masses (the only way to measure BH masses at high
redshift).\\ {\bf b)} constrain  geometry and kinematics of the BLR.\\
{\bf c)}  if BLR is in  a rotating disk,  the BH mass can  be directly
measured.

\begin{figure}
\centering
\includegraphics[angle=90,width=1.0\textwidth]{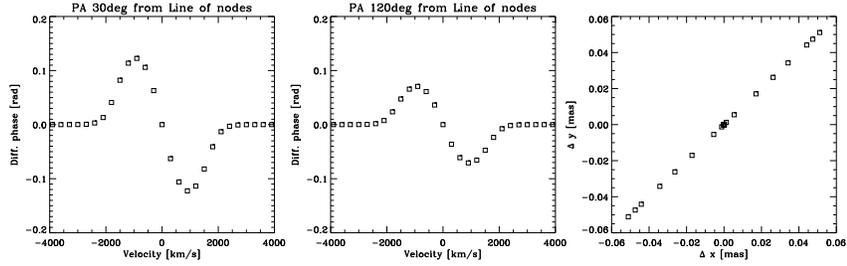}
\caption{\label{fig:blr}  Observations of  the BLR  of 3C273  which is
  assumed to  be a  gas disk inclined  of 30  deg w.r.t.  the  line of
  sight and rotating around a  $3\times 10^8\ M_\odot$ BH. In the left
  and middle  panel we show  the expected line differential  phase per
  wavelength bin  along 2  possible baselines. In  the right  panel we
  show  the  reconstructed positions  of  the  line  centroid in  each
  wavelength bin.  The line centroids  are aligned along the disk line
  of nodes  combining their  relative positions with  their associated
  velocity it is possible to directly measure the BH mass.  }
\end{figure}

\section{Supermassive blackholes}

To detect a BH one must  resolve the radius of its sphere of influence
which, projected on the plane of the sky, is
\begin{eqnarray} 
\theta_{BH}=0.1\prime\prime\left(\frac{M_{\rm BH}}{10^7 M_\odot}\right)
\left(\frac{\sigma_\star}{100\ {\rm km/s}}\right)^{-2} 
\left(\frac{D}{10\ {\rm Mpc}}\right)^{-1}
\end{eqnarray}
It is  clear that in  order to probe  the BH sphere of  influence high
spatial resolution is  mandatory. Until now, except for  the Milky Way
and NGC 4258, the best  spatial resolution achievable is given by HST,
i.e.   $\sim 0.1\prime\prime$.

\begin{figure}
\centering \includegraphics[angle=90,width=0.9\textwidth]{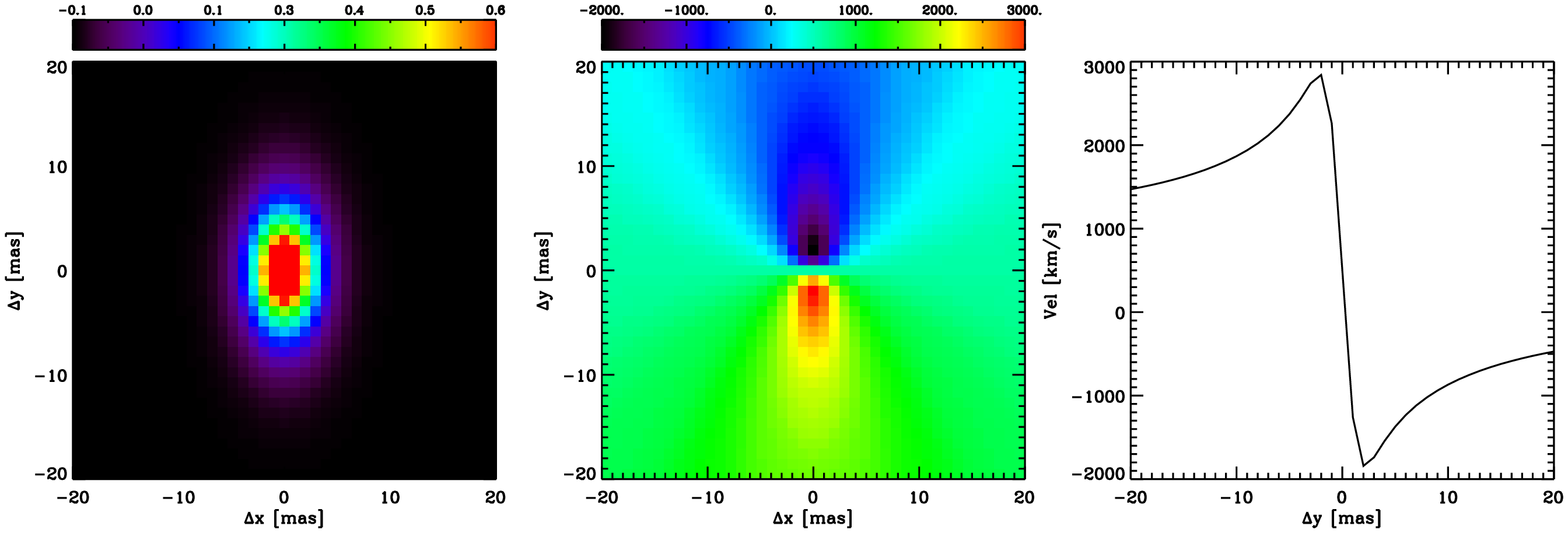}
\includegraphics[angle=90,width=0.9\textwidth]{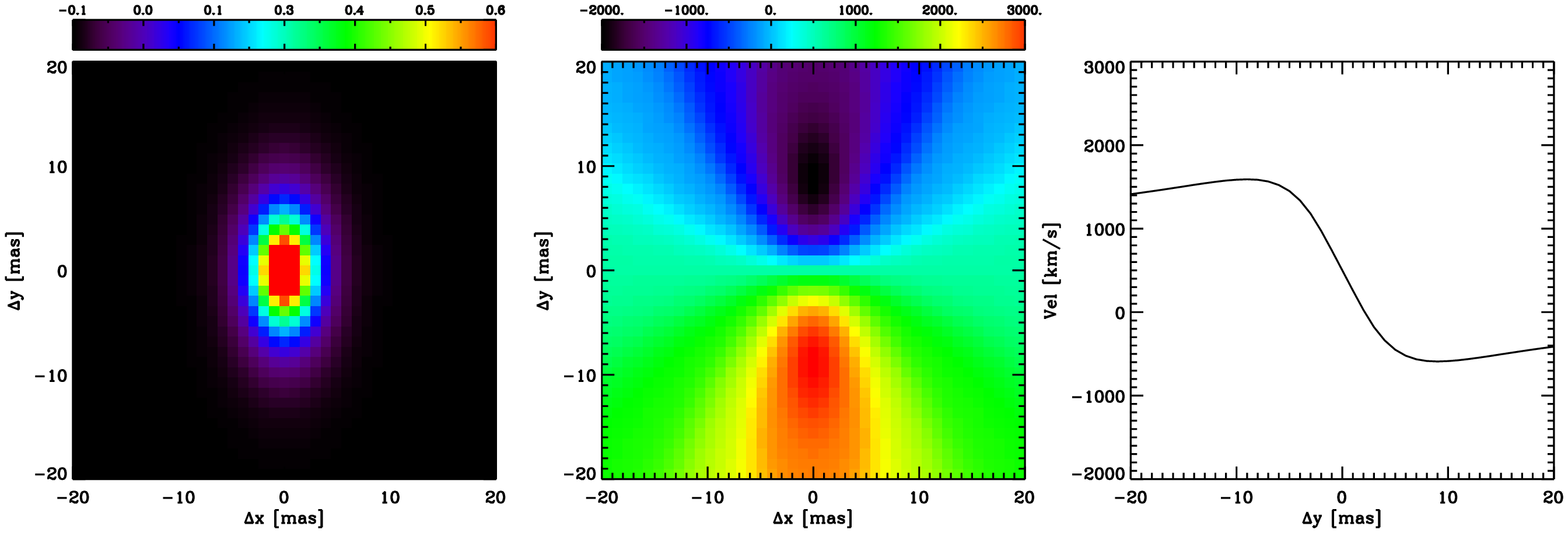}
\caption{\label{fig:cena} Left panels:  Line images (e.g.\ Br$\gamma$)
  of the rotating gas disk of  Centaurus A which is inclined by 60 deg
  w.r.t.  the line of sight and has the major axis along the $y$ axis.
  The line images have pixel sizes  of 1 mas and spatial resolution of
  2 mas. Middle panels: velocity  fields of the rotating gas disk with
  the same pixel size and  spatial resolution as before.  Left panels:
  velocity curves along the major axis.  Top panels refers to the case
  of  a central  supermassive  BH of  $10^8\  M_\odot$, while  bottom
  panels consider  the case of  an extended massive dark  objects with
  $10^8\ M_\odot$ mass  and core radius of 5 mas.   The two cases are
  clearly distinguishable kinematically.  }
\end{figure}

{\bf Galactic Center}  VITRUV will allow to measure  proper motions of
stars  near the  Galactic Center  BH with  an accuracy  better  than 1
mas.\\ {\bf a)}  It will be possible to  further constrain the spatial
extent of  the dark  mass concentration (the  supposed BH)  to exclude
even the  last possible  alternative, the boson  star.\\ {\bf  b)} The
knowledge of orbits with such a high accuracy will allow to search for
general  relativistic effects like  the periastron  precession.\\ {\bf
  c)} It will be possible  to disentangle the flaring IR emission near
the BH  event horizon from the  background emission.  Thus  it will be
possible to obtain accurate light curves and determine the periodicity
of the emission which might lead to a measurement of the BH spin.

{\bf Extragalactic  BHs} Combining  the imaging and  the spectroscopic
capabilities  at medium resolution  ($>1000$) it  will be  possible to
obtain data cubes of emission lines around the central BHs.  This will
allow to study the morphology and kinematics of the ionized gas.  From
these studies it will be possible to:\\
{\bf a)} constrain the size of the massive dark objects in nearby
galactic nuclei to exclude plausible alternatives to BHs;\\
{\bf b)} directly measure the BH mass up to a distance which is a factor
$\sim 50$ larger than possible nowadays.\\
{\bf c)} establish ''secure'' BH mass vs hot galaxy properties relations.\\
Fig.~\ref{fig:cena}  clearly show  how it  is possible  to distinguish
between the cases of a supermassive  BH or of an extended massive dark
cluster.

\section{Conclusion}

In these paper  we presented a series of science  cases for the VITRUV
second generation  VLTI instrument concept.  The focus  was on imaging
as this  is the novel routine  capability enabled by  this instrument. 
VITRUV is a general purpose  instrument. The science is not exhaustive
but  illustrative, and  complementary to  the PRIMA  reference missions
(Perrin et al. 2004). The following areas
were addressed:\\
{\bf a)} the close  environment of young stars;\\
{\bf b)} activity in the surfaces of stars;\\
{\bf c)} the  origins of Planetary Nebulae geometries;\\
{\bf d)} accretion-ejection structures  in micro-quasar;\\
{\bf e)} the components (torus, jets,  BLRs) of AGNs;\\
{\bf  f)}  the  environment  of nearby  supermassive  black-holes;\\
{\bf g)} relativistic effects in the Galactic Center black-hole.

These cases  together with the  dramatic operational gains  allowed by
VITRUV  clearly   underline  the  importance  of   a  general  purpose
spectro-imager for the second generation VLTI instrument.

\begin{acknowledgement}
  PJVG  work  was supported  in  part  by  the Funda\c{c}\~ao  para  a
  Ci\^encia  e a  Tecnologia through  project POCTI/CTE-AST/55691/2004
  from POCTI, with funds from the European programme FEDER.
\end{acknowledgement}



\printindex
\end{document}